\documentclass[aps,twocolumn]{revtex4}
\usepackage{amsmath,amssymb}

\usepackage{epsfig}
\usepackage{graphicx}

\newcommand{\be}{\begin{equation}}
\newcommand{\ee}{\end{equation}}
\newcommand{\bea}{\begin{eqnarray}}
\newcommand{\eea}{\end{eqnarray}}
\newcommand{\nn}{\nonumber}

\begin{document}

\title{Cosmology with exponential potentials}

\author{Alex Kehagias$^{1,2}$\footnote{kehagias@mail.cern.ch} and Georgios Kofinas$^{1,3}$\footnote{kofinas@cecs.cl}}

\date{\today}


\address{$^{1}$ Department of Physics, National Technical
University of Athens, GR~157~73~Athens, Greece}

\address{$^{2}$ Institute of Nuclear Physics, N.C.S.R. Democritos, GR~153~10~Athens, Greece}

\address{$^{3}$ Centro de Estudios Cient\'{\i}ficos,
Casilla~1469, Valdivia, Chile}

\begin{abstract}

We examine in the context of general relativity the dynamics of a
spatially flat Robertson-Walker universe filled with a classical
minimally coupled scalar field $\phi$ of exponential potential
$V(\phi)\sim exp(\!-\mu\phi)$ plus pressureless baryonic matter.
This system is reduced to a first-order ordinary differential
equation for $\Omega_{\phi}(w_{\phi})$ or $q(w_{\phi})$, providing
direct evidence on the acceleration/deceleration properties of the
system. As a consequence, for positive potentials, passage into
acceleration not at late times is generically a feature of the
system for any value of $\mu$, even when the late-times attractors
are decelerating. Furthermore, the structure formation bound,
together with the constraints $\Omega_{m0}\approx 0.25-0.3$,
$-1\leq w_{\phi 0}\leq -0.6$ provide, independently of initial
conditions and other parameters, the necessary condition
$0<\mu\lesssim 1.6\sqrt{8\pi G_{N}}$, while the less conservative
constraint $-1\leq w_{\phi}\leq -0.93$ gives $0<\mu\lesssim
0.7\sqrt{8\pi G_{N}}$. Special solutions are found to possess
intervals of acceleration. For the almost cosmological constant
case $w_{\phi}\approx -1$, the general relation
$\Omega_{\phi}(w_{\phi})$ is obtained. The generic
(non-linearized) late-times solution of the system in the plane
$(w_{\phi}, \Omega_{\phi})$ or $(w_{\phi}, q)$ is also derived.

\end{abstract}

\maketitle
\section{Introduction}

Scalar-field cosmologies continuously receive through time special
attention. The interest on scalar fields has been stimulated when
it has been realized that a scalar field might be responsible for
inflation of the very early universe. If its self-interaction
potential energy density is sufficiently flat, it can violate the
strong energy condition and drive the universe into an accelerated
expansion. Even if the potential is too steep to drive inflation,
such models still have important cosmological consequences. Recent
observational evidence shows that the energy density of the
universe today may be dominated by a homogeneous component with
negative pressure (quintessence)~\cite{perl}. Slowly rolling
self-interacting scalar fields may provide in the late-times
evolution of the universe a dynamical mechanism to achieve a small
effective cosmological constant.
\par
Models with a variety of self-interaction potentials have been
studied and the cosmological conclusions are naturally strongly
model dependent. There are attempts to base the choice of the
potential on some fundamental dynamical considerations. A class of
potentials that is commonly investigated and which arises in a
number of physical situations has an exponential dependence on the
scalar field. Such potentials are motivated by higher-dimensional
supergravity, superstring theory, or M-theory  that have been
compactified to an effective four-dimensional theory~\cite{sugra}.
Also, higher-order gravity theories are conformally equivalent to
general relativity plus a scalar field with potential which
asymptotically tends to an exponential form~\cite{conformal}.
\par
Since the first inflationary cosmological models were presented,
most calculations have been performed using one of a variety of
approximations. While these are usually sufficient and often
unavoidable, there is considerable interest in scalar field
cosmologies that can be solved exactly. Exponential potentials have
been extensively considered in cosmology by various authors,
basically in Robertson-Walker models with no other source besides
the scalar field~\cite{cosmology}. Solutions with
anisotropic/inhomogeneous geometries were obtained
in~\cite{anisotropic}. When a perfect fluid is added, only a few
solution are known (see references~\cite{matter}-\cite{heard} for
relevant works). In the context of ekpyrotic or pre-big bang
collapse, exponential potentials were analyzed in~\cite{ekpyrotic},
while non-minimal couplings of scalar fields with exponential
potentials have been considered in~\cite{nonminimal}. Attention has
also been directed to the possible effect of such a scalar field on
the growth of large-scale structure in the
universe~\cite{structure}. Other recent studies on exponential
potentials have appeared in~\cite{recent}.
\par
In the present paper, we examine the dynamics of a spatially flat
Robertson-Walker universe filled with a classical minimally
coupled scalar field of exponential potential plus pressureless
baryonic matter (luminous and nonluminous) non-interacting with
the field. Clearly, this model cannot be used from the beginning
of the universe evolution, but only after decoupling of radiation
and dust. Thus, we do not take into account inflation, creation of
matter, nucleosynthesis, e.t.c. By performing changes of
phase-space variables, we reduce the system of Einstein equations
into a first order nonlinear differential equation, which can be
expressed as an equation for $\Omega_{\phi}(w_{\phi})$ or
$q(w_{\phi})$. Solving this equation leads to integration of the
whole system. Furthermore, studying this with respect to its
acceleration properties, as well as with respect to standard
observational constraints, restrictions on the parameter
determining the steepness of the potential are obtained. For the
almost cosmological constant case $w_{\phi}\approx -1$, the
solution $\Omega_{\phi}(w_{\phi})$ is obtained. The late-times
behaviour $\Omega_{\phi}(w_{\phi})$ of the system is also found.

\section{General analysis of the model}

We consider the action \bea  && S=\frac{1}{2\kappa^{2}}\int
R\sqrt{-g}\,d^{4}x + \int L_{mat}\sqrt{-g}\,d^{4}x\nn\\
 &&~~~~~~~{}-\int
\Big(\frac{1}{2}g^{\mu\nu}\partial_{\mu}\phi\partial_{\nu}\phi+V\Big)\sqrt{-g}\,d^{4}x
\,,  \label{action} \eea with the term $L_{mat}$ corresponding to
the perfect fluid, and we adopt the Robertson-Walker metric \be
ds^{2}=-dt^{2}+a^{2}(t)\gamma_{ij}dx^{i}dx^{j}\,,
\label{lineelement}
 \ee where
$\gamma_{ij}$ is a maximally symmetric 3-dimensional metric, and
$k=-1,0,1$ parametrizes the spatial curvature. The Einstein
equations for the system of a perfect fluid and a scalar field is
equivalent to the following set of equations \be
\dot{\rho}+3\frac{\dot{a}}{a}(\rho+p)=0 \label{rhoconserv} \ee \be
\dot{\rho}_{\phi}+6\frac{\dot{a}}{a}(\rho_{\phi}-V)=0
\label{rhophiconserv} \ee \be
H^{2}=\frac{\dot{a}^{2}}{a^{2}}=\beta(\rho+\rho_{\phi})-\frac{k}{a^{2}}\,,
\label{hubble} \ee where $\beta=\kappa^{2}/3=8\pi G_{N}/3$, a dot
means differentiation with respect to proper time $t$, and
$\rho_{\phi}$ is the energy density of the scalar field defined by
\be \rho_{\phi}=\frac{1}{2}\dot{\phi}^{2}+V(\phi)\,.
\label{rhophi} \ee Since the perfect fluid and the scalar field
are non-interacting, each separately satisfies its own
conservation equation (\ref{rhoconserv}), (\ref{rhophiconserv})
respectively. Equation (\ref{rhophiconserv}) is equivalent to the
classical equation of motion $\bigtriangleup \phi=V'(\phi)$ of the
field $\phi$, where $\bigtriangleup$ is the 4-dimensional
Laplacian of the metric (\ref{lineelement}) and a prime denotes
differentiation with respect to $\phi$. From
Eq.~(\ref{rhophiconserv}), it arises that the function
$\rho_{\phi}(t)$ is monotonically decreasing (increasing) for an
expanding (contracting) universe.
\par
From Eqs.~(\ref{rhoconserv}), (\ref{rhophiconserv}),
(\ref{hubble}), one can derive the Raychaudhuri equation \be
\dot{H}=-\frac{3\beta}{2}(\rho+p+\dot{\phi}^{2})+\frac{k}{a^2}\,,
\label{ray} \ee which is also written in the form \be
\frac{\ddot{a}}{a}=\beta
\Big(3V-2\rho_{\phi}-\frac{1+3w}{2}\rho\Big)\,, \label{accel} \ee
where $w=p/\rho$ characterizes the perfect fluid.
\par
Defining the quantity \be \varrho = \rho^{2}+\rho_{\phi}\,,
\label{rhoitalic} \ee we obtain from Eqs.~(\ref{rhoconserv}),
(\ref{rhophiconserv}) that \be
\dot{\varrho}+6\frac{\dot{a}}{a}(\varrho-V+\rho p)=0\,.
\label{rhoitalicconserv} \ee We will specify to a flat universe
($k=0$) with dust matter ($w=0$). Equation
(\ref{rhoitalicconserv}), due to (\ref{hubble}), is written for an
expanding universe as \be \sqrt{\rho_{\phi}-V}\,\varrho'=\mp
3\sqrt{2\beta}\,\left(\sqrt{\varrho-\rho_{\phi}}+\rho_{\phi}\right)^{1/2}(\varrho-V)\,,
\label{rhoitalicprime} \ee where $\dot{\phi}=\pm
\sqrt{2}\sqrt{\rho_{\phi}-V}$. Similarly,
Eq.~(\ref{rhophiconserv}) takes the form \be \rho_{\phi}'=\mp
3\sqrt{2\beta}\,\left(\sqrt{\varrho-\rho_{\phi}}+\rho_{\phi}\right)^{1/2}\sqrt{\rho_{\phi}-V}.
\label{rhophiprime} \ee Dividing Eqs.~(\ref{rhoitalicprime}) and
(\ref{rhophiprime}), we get \be
\frac{\rho_{\phi}'}{\varrho'}=\frac{\rho_{\phi}-V}{\varrho -V}\,.
\label{rhofraction} \ee The system of
Eqs.~(\ref{rhoconserv})-(\ref{hubble}) is symmetric under time
reversal $t\rightarrow -t$, thus a contracting universe arises
from an expanding one. Additionally, the stable fixed points of
the system for an expanding universe become unstable fixed points
in a contracting universe and vice-versa, thus we do not discuss
the collapsing behaviour further.
\par
We consider a potential with an exponential-like form \be
V(\phi)=V_{0}e^{-\mu\phi}\,, \label{potential} \ee and define \be
R=\frac{\rho_{\phi}}{V}\,\,\,\,\,,\,\,\,\,\,Q=\frac{\sqrt{\varrho
-\rho_{\phi} }}{V}\,. \label{RQ} \ee The system of
Eqs.~(\ref{rhophiprime}), (\ref{rhofraction}) is written
equivalently as \be Q=\frac{(R'-\mu R)^{2}}{18\beta(R-1)}-R
\label{Q} \ee \be (\ln Q)'=\frac{R'+\mu R-2\mu}{2(R-1)}\,,
\label{Qprime} \ee under the conditions $\pm V_{0}(R'-\mu R)<0$,
$(R'-\mu R)^2>18\beta R(R-1)$. One characteristic of this system
is that, due to the exponential potential, it does not contain the
independent variable $\phi$ explicitly. Since the system of
Eqs.~(\ref{rhoconserv})-(\ref{hubble}) is symmetric under the
simultaneous change $\mu\rightarrow -\mu$, $\phi \rightarrow
-\phi$, we will consider, without loss of generality, only $\mu >
0$. Combining Eqs.~(\ref{Q}), (\ref{Qprime}), we can eliminate
$Q$, and obtain \be R'-\mu R=0 \label{axristi} \ee or \bea &&
4(R-1)R''-3R'^{2}-2\mu(R-3)R'\nn
\\&&~~~~~~~~~~{} -(R-2)\left((18\beta-\mu^{2})R-18\beta\right)=0\,. \label{Rdoubleprime} \eea
Equation (\ref{axristi}) is equivalent to $\rho_{\phi}=-\rho$,
which gives from Eq.~(\ref{hubble}) a static universe $a=$
constant, and is not of further interest here. Making the
transformation \be
\chi=R-1\,\,\,\,\,,\,\,\,\,\,\psi=\frac{\mu}{R'}\,, \label{xy}\ee
the second order differential equation (\ref{Rdoubleprime})
reduces to the following first order differential equation in
$\psi(\chi)$: \be
4\chi\frac{d\psi}{d\chi}=(\chi-1)(1-\nu^{-1}\chi)\psi^{3}-2
(\chi-2)\psi^{2}-3\psi\,, \label{dydx} \ee where
$\nu^{-1}+1=18\beta/\mu^{2}$. In terms of $\chi, \psi$, the
previous constraints take the form $\pm
V_{0}(\psi^{-1}\!-\!\chi\!-\!1)<0$,
$(\psi^{-1}\!-\!\chi\!-\!1)^{2}>(\nu^{-1}\!+1)\chi(\chi+1)$.
\newline
In terms of the original variables, it is \be
\chi=\frac{\dot{\phi}^{2}}{2V}\,\,\,\,\,,\,\,\,\,\,\psi=\mu\Big(\frac{d\chi}{d\phi}\Big)^{-1}\,,
\label{original} \ee and thus $V_{0}\chi>0$. Finding solutions of
Eq.~(\ref{dydx}) is the starting point for running the equations
backwards to find solutions of the whole system. Equation
(\ref{dydx}) belongs to the class of Abel equations. There is no
known systematic way for solving such sort of equations. For the
same problem with the one examined here, another Abel equation,
inequivalent to (\ref{dydx}), was obtained in Ref.~\cite{abel}.
The difference is that there, a transformation of the
configuration variables has been performed.
\par
Given a solution $\psi(\chi)$ of Eq.~(\ref{dydx}), one can find
all the relevant physical quantities in parametric form. It is
obvious that $\chi$ is not in general a global time parameter, but
it remains a good time parameter, as long as $\chi(t)$ is
monotonically increasing or decreasing. So, the scalar field
$\phi(\chi)$ is found from Eq.~(\ref{original}) as \be
\phi(\chi)=\phi_{1}+\mu^{-1}\!\int{\!\psi(\chi)d\chi}\,,
\label{find} \ee where $\phi_{1}$ is integration constant. Then,
the scale factor (normalized to unity today) is expressed as \be
a^{3}(\chi)=\frac{1}{V_{0}}\frac{\rho_{0}(\nu^{-1}\!+\!1)\chi\,e^{\mu\phi(\chi)}}
{(\psi^{-1}\!-\!\chi\!-\!1)^{2}\!-(\nu^{-1}\!+\!1)\chi(\chi+1)}\,,
\label{sca} \ee with $\rho_{0}$ the present value of $\rho$, and
the Hubble parameter is \be H^{2}(\chi)=\frac{\mu^2
V_{0}}{18\chi}e^{-\mu\phi(\chi)}(\psi^{-1}\!-\!\chi\!-\!1)^{2}\,.
\label{hu} \ee The luminosity distance
$d_{L}\!=\!(1+z)\int_{t}^{t_{0}}{\!dt/a}$, which is directly
measured, is given by \bea
\!\!\!\!\!\!\!\!\!d_{L}(z)&\!\!=\!\!&\pm
\frac{1+z}{\sqrt{2}\mu\rho_{0}^{1/3}}
\!\times\nn\\
&&\!\!\!\!\!\!\!\!\!\!\!\!\!\!\!\!\!
\int_{\chi}^{\chi_{0}}\!\!{\frac{\psi(\chi)}{\sqrt{V_{0}\chi}}e^{\frac{1}{6}\mu\phi(\chi)}
\Big{\{}\!V_{0}\Big[\frac{(\psi^{-1}\!-\!\chi\!-\!1)^2}{(\nu^{-1}\!+\!1)\chi}\!-\!\chi\!-\!1\Big]\!\Big{\}}
^{\!\frac{1}{3}}d\chi}, \label{lum} \eea where the redshift $z$ is
defined by $1+z=1/a$, and $\chi_{0}$ denotes the value of $\chi$
today. The flatness parameters $\Omega_{m}=\beta\rho/H^2$,
$\Omega_{\phi}=\beta\rho_{\phi}/H^2$ are found as \be
\Omega_{\phi}(\chi)=\frac{(\nu^{-1}+1)\chi(\chi+1)}{(\psi^{-1}-\chi-1)^{2}}=1-\Omega_{m}(\chi)\,.
\label{fla} \ee Similarly, the energy density and the pressure
$p_{\phi}=(\dot{\phi}^{2}/2)-V$ of the scalar field are given
respectively by
$\rho_{\phi}(\chi)=V_{0}(\chi+1)e^{-\mu\phi(\chi)}$,
$p_{\phi}(\chi)=V_{0}(\chi-1)e^{-\mu\phi(\chi)}$. Besides these
parametric expressions, a physically interesting point is the
determination of the equation of state
$p_{\phi}=p_{\phi}(\rho_{\phi})$ of the scalar field, or
equivalently of the function $w_{\phi}(\rho_{\phi})$, where
$w_{\phi}=p_{\phi}/\rho_{\phi}$ is the state parameter of the
scalar field, given by \be
w_{\phi}=\frac{\chi-1}{\chi+1}\,\Leftrightarrow\,\chi=\frac{1+w_{\phi}}{1-w_{\phi}}.
\label{chiwphi} \ee The inequality $V_{0}\chi>0$ translates to
$|w_{\phi}|<1$ for $V_{0}>0$, and $|w_{\phi}|>1$ for $V_{0}<0$.
Then, it is found that \be
\rho_{\phi}(w_{\phi})=\frac{2V_{0}}{e^{\mu\phi_{1}}}\frac{1}{1-w_{\phi}}
e^{-2\int{\frac{\psi(\chi(w_{\phi}))}{(1-w_{\phi})^2}}\,dw_{\phi}}\,\,.
\label{twra} \ee In this language, Eq.~(\ref{dydx}) becomes a
second order differential equation for
$\ln{\rho_{\phi}}(w_{\phi})$. Concerning the problem of
acceleration, we can find the decelerating parameter
$q=-\ddot{a}/aH^{2}$ from Eqs.~(\ref{accel}), (\ref{rhoitalic}),
(\ref{RQ}), (\ref{Q}) as \be
q=\frac{1}{2}-\frac{3(\nu^{-1}+1)\chi(1-\chi)}{2(\psi^{-1}-\chi-1)^{2}}=
\frac{1}{2}(1+3w_{\phi}\Omega_{\phi})\,, \label{q}\ee Since the
variables $\chi, \psi$ are phase space variables, by studying
Eq.~(\ref{dydx}), it is straightforward from Eqs.~(\ref{q}) to get
information on the acceleration/deceleration properties of the
cosmological system. It is seen from the first equality of
Eq.~(\ref{q}) that acceleration is possible only for
$-1<w_{\phi}<0$, and thus, it is possible only for $V_{0}>0$.
Additionally, for $V_{0}>0$, since $0<\Omega_{\phi}<1$, it arises
from the second equality of Eq.~(\ref{q}) that acceleration is
possible only for $-1<w_{\phi}<-1/3$. This is an exact result,
without the usual assumption on the dominance of the quintessence
component in the energy density, and of course, it is only a
necessary condition. Finally, the relation between $\chi$ and $t$
is found by \be
t(\chi)=t_{1}\pm\frac{1}{\mu\sqrt{2}}\int{\frac{\psi(\chi)}{\sqrt{V_{0}\chi}}}\,e^{\mu\phi(\chi)/2}\,d\chi\,,
\label{ti} \ee with $t_{1}$ integration constant. We observe that
the number of essential integration constants of the system is
three, namely $\rho_{0}, \phi_{1}$, and the one contained in
$\psi(\chi)$ from integration of  Eq.~(\ref{dydx}). This is also
predicted from the beginning, since the system consists of two
first-order equations (\ref{rhoconserv}), (\ref{hubble}), and one
second-order equation (\ref{rhophiconserv}), giving four
integration constants, from where one of them is taken out due to
the time translation.
\par
In Ref.~\cite{fixed}, for $V_{0}>0$ and \be
x=\sqrt{\frac{\beta}{2}}\frac{\dot{\phi}}{H}\,\,\,\,\,\,\,\,,\,\,\,\,\,\,
y=\frac{\sqrt{\beta\,V}}{H}\label{mum}\,, \ee the system of
Eqs.~(\ref{rhoconserv})-(\ref{hubble}) was written as a
plane-autonomous system \bea
&&\frac{dx}{d\ln{a}}=-3x+\frac{\mu}{\sqrt{2\beta}}y^{2}+\frac{3}{2}x(1+x^{2}-y^{2})\label{fou}\\
&&
\frac{dy}{d\ln{a}}=-\frac{\mu}{\sqrt{2\beta}}xy+\frac{3}{2}y(1+x^{2}-y^{2})\label{ftou}\,.\eea
It was further found that for any value of $\mu$, there are three
kinds of fixed points, characterized by the values of $(x, y)$:
(i) $(\pm 1,\,0)$ , (ii) $(0,\,0)$ , (iii) $(\mu,\,
\sqrt{18\beta-\mu^{2}})/3\sqrt{2\beta}$ for $\mu^2 < 18\beta$ and
$3\sqrt{\beta}(1,\,1)/\sqrt{2}\mu$ for $\mu^2 > 9\beta$. The fixed
points (i) are unstable nodes or saddle points, the fixed point
(ii) is a saddle point, while the fixed points (iii) are stable
points or saddles. More specifically, for any value of $\mu$
($\mu^{2}\neq 9\beta, 72\beta/7$) there exists one stable fixed
point, which is accelerating for $\mu^2<6\beta$, and decelerating
for $\mu^2>6\beta$. For $V_{0}<0$, generic solutions recollapse to
a singularity after a finite lifetime \cite{heard}. The relation
between the variables $x, y$ and $\chi, \psi$ can be found from
the previous definitions as \bea
\chi=x^{2}/y^{2}\,\,\,\,\,\,,\,\,\,\,\,\,\psi^{-1}=1+x(x-3\sqrt{2\beta}/\mu)y^{-2}\label{pop}\,.
\eea

\section{Cosmological implications}

We are not in position to find the generic solution of
Eq.~(\ref{dydx}), and then, study from Eq.~(\ref{q}) the
accelerating properties of the system. However, if
$f=\Omega_{m}/\Omega_{\phi}=(1/\Omega_{\phi})-1$ is the fraction
of the flatness parameters, we are able to study the accelerating
properties of the system by deriving from Eqs.~(\ref{dydx}),
(\ref{q}) the following first order differential equation for the
function $f(w_{\phi})$ \bea
&&\!\!\!\!\!\!\!\!\!\frac{w_{\phi}^2\!-\!1}{w_{\phi}(f+1)}\frac{df}{dw_{\phi}}=
1\mp
\frac{\epsilon}{\sqrt{(\nu^{-1}+1)(w_{\phi}+1)(f+1)}}\times\nn\\
&&\,\,\,\,\,\,\,\,\,\,\,\,\,\,\,\,
\Bigg[\frac{(1-\nu^{-1})-(1+\nu^{-1})w_{\phi}}{\sqrt{2}\!\mp\!\epsilon
\sqrt{(\nu^{-1}\!+\!1)(w_{\phi}+1)(f+1)}}-1\Bigg],
\label{form}\eea
 where $\epsilon=sgn(V_{0}(1-w_{\phi}))$, and then,
 $2q(w_{\phi})=1+3w_{\phi}/(1+f(w_{\phi}))$. For $V_{0}>0$, due to the above discussion on the
fixed points, it is obvious that all the solutions of the system
start their evolution with deceleration at $\chi=+\infty$
($w_{\phi}=1$). Concerning the parameter $\chi$, we can obtain its
time evolution as
$\dot{\chi}=\dot{\phi}(\mu\rho_{\phi}-3H\dot{\phi})/V=\pm
\mu\sqrt{2V\chi}\,\psi^{-1}$, thus, for $V_{0}>0$ and
$\dot{\phi}<0$, the function $\chi(t)$ (and also the function
$w_{\phi}(t)$) is monotonically decreasing. This means that all
the solutions emanating from the repeller $(-1,\,0)$ have an
initial period, characterized by $\dot{\phi}<0$, where $\chi(t)$
is monotonically decreasing from $\chi=+\infty$ to $\chi=0$ (and
also $w_{\phi}$ monotonically decreases from $w_{\phi}=1$ to
$w_{\phi}=-1$). We have plotted the solutions of Eq.~(\ref{form})
for these initial time intervals in Figure 1, shown by the family
of ``circular'' lines, where an indicative value of the parameter
$\nu=3/4$ has been used. The deceleration parameter $q(w_{\phi})$
is shown in the same figure by the inclined lines passing through
the point $1/2$ of the vertical axis, providing direct evidence on
the acceleration/deceleration properties of the system. Lines of
the same indication in Figure 1 correspond to the same solution,
while time increases from the right to the left along the curves.
Using a first-order differential equation, since only one initial
value enters, there is the merit of representing the whole space
of solutions on a plane diagram. It is seen that there is a
subclass of solutions which enter from initial deceleration to an
acceleration era. However, there are still solutions with
$w_{\phi}<-1/3$ \,which remain decelerated. It is easy to check
from Eq.~(\ref{form}) that the structure of Figure 1 is typical
for any value of $\nu$. Thus, even for $\nu>1/2$, where the
late-times attractor is decelerating, there are certain classes of
solutions which have finite accelerating intervals not at late
times. Actually, the supernovae data favour that the present-day
accelerated expansion is a recent phenomenon $z_{q=0}\approx 0.5$
\cite{turner}. Nucleosynthesis predictions \cite{nucle}, on the
other hand, claim that at 95 per cent confidence level, it is
$\Omega_{\phi}(z= 10^{10})\leq 0.045$. Besides this, during galaxy
formation epoch ($z=2-4$), the quintessence density parameter is
$\Omega_{\phi}<0.05$ \cite{tur}. This is an additional constraint,
applicable in the dust regime, which restricts the viable
solutions. For instance, for the indicative value $\nu=3/4$ of
Figure 1, no accelerating solution can accommodate this bound,
which means that the present accelerating era, if possible, has to
be located in some future time having $\dot{\phi}>0$. For
different values of $\nu$, however, the above bound may match with
the present-day acceleration, already in the epoch with
$\dot{\phi}<0$.
\begin{figure}[h!]
\centering
\includegraphics*[width=189pt, height=140pt]{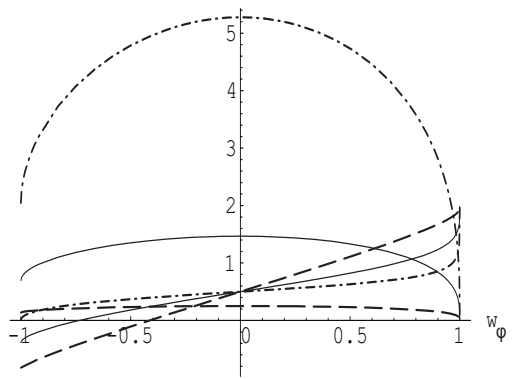}%
\caption{The class of all solutions with $V_{0}>0$, $\dot{\phi}<0$
for an indicative value of $\nu=3/4$. Family of ``circular'' lines
represent $(\Omega_{m}/\Omega_{\phi})(w_{\phi})$, while inclined
lines passing through the point $1/2$ of the vertical axis
represent $q(w_{\phi})$. Lines of the same indication correspond
to the same solution. The scale factor increases from the right to
the left along the curves. A subclass of solutions show passage
from deceleration to acceleration not at late times. Similar
portraits occur for all values of $\nu$.}
\end{figure}
\par
In order to pass into the region with $\dot{\phi}>0$, an
appropriate time variable, covering a region with both signs of
$\dot{\phi}$, is the parameter $\Omega_{m}$, or equally well
$\Omega_{\phi}$, or also $f$ (we are going to use $f$). This is
due to that, in general, $\dot{\Omega}_{m}=3\beta^2 \rho
V(\chi-1)/H^3$, and thus, for $V_{0}>0$, $\chi<1$ (or
$-1<w_{\phi}<0$), the functions $\Omega_{m}(t), f(t)$ are
monotonically decreasing. Equation (\ref{form}) can, then, be
easily inverted, and considered as a first order differential
equation for $w_{\phi}(f)$, and then,
$2q(f)=1+3w_{\phi}(f)/(1+f)$. The solutions $w_{\phi}(f)$
(``lower'' curves of the same indication), $q(f)$ (``upper''
curves) are shown in Figure 2 for the value $\nu=3/4$, where time
increases from the right to the left along the curves. Lines of
the same indication in Figure 2 correspond to the same solution,
and also, they are the continuations of the corresponding curves
of Figure 1.
\begin{figure}[h!]
\centering
\includegraphics*[width=169pt, height=130pt]{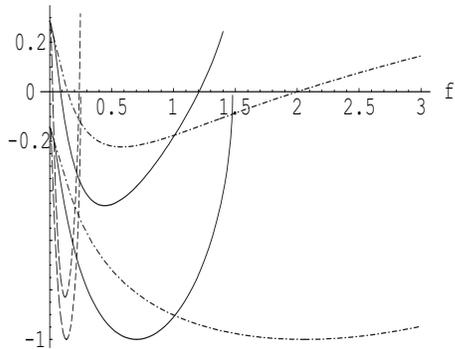}%
\caption{Class of solutions with $V_{0}>0$, $-1<w_{\phi}<0$ for an
indicative value of $\nu=3/4$. ``Lower'' lines of the same
indication represent $w_{\phi}(f)$, while ``upper'' ones represent
$q(f)$. Lines of the same indication correspond to the same
solution, and they are the continuations of the corresponding
curves of Figure 1. The scale factor increases from the right to
the left along the curves. Similar portraits occur for all values
of $\nu$.}
\end{figure}
It is well known that a set of complementary observations indicate
that the matter energy density $\Omega_{m0}$ is no more than $0.3$
\cite{wang}, while observations of the higher peaks of the cosmic
background radiation give $\Omega_{m0}=0.25$ \cite{cora}. A
conservative estimation for the present quintessence equation of
state is $-1\leq w_{\phi 0}\leq -0.6$ \cite{wang}. It is now a
simple task, using figures similar to Figures 1 and 2, to
determine for any particular value of $\nu$, if there are
accelerating solutions satisfying these two constraints on
$\Omega_{m0}, w_{\phi 0}$, together with the previous one from
structure formation. The result of this analysis is that the above
bounds are satisfied only for $\nu \lesssim 3/4$, with agreement
to previous studies on exponential potentials, based on numerical
analysis of the full second-order equations of the system
\cite{kolda}. This range of $\nu$ corresponds to a scalar field
dominated attractor ($\Omega_{\phi}=1$), with limiting value
$w_{\phi}=(1-\nu^{-1})/(1+\nu^{-1})$. Although there is no
explicit dependence on the scale factor $a$ or the redshift $z$ in
Eq.~(\ref{form}), and therefore in Figures 1 and 2, however, in
the previous analysis there is the merit of providing direct
information on the viability of whole spaces of solutions, without
particular reference to integration constants or other parameters
of the system. Additionally, the non-dependence of $w_{\phi}$ on
$z$ is profitable for a quantitative treatment, as it is known
that $w_{\phi}$ changes very rapidly between the redshifts of
various supernovae data and the present.

\subsection{Approximate solution for $w_{\phi}\approx -1$}

The values $w_{\phi}\approx -1$ are of particular interest, since
firstly, it is known \cite{kolda,ena} to cover various orders of
magnitude in scale factor during the cosmic evolution, and
secondly, for a class of potentials larger than those examined
here, a combined analysis \cite{cora} including all three peaks in
the Boomerang data of the CMBR spectra, and the Type Ia
supernovae, gave as best fit $\Omega_{\phi 0}=0.75$, $-1\leq
w_{\phi 0}\leq -0.93$, thus, behaviour similar to a cosmological
constant. Considering this restrictive bound for $w_{\phi 0}$, and
following the previous analysis, we are restricted to $\nu\lesssim
0.08$ as necessary condition for the viability of the exponential
model. In this case, we are in position to find the relation
$\Omega_{\phi}(w_{\phi})$. Making the transformation \be
\xi=\eta\chi^{-1}\,\,\,\,,\,\,\,\,\Upsilon=(\eta\chi)^{-3/4}\psi^{-1}\,,
\label{transform} \ee where $\eta= sgn(V_{0})$, Eq.~(\ref{dydx})
takes the form \be
\Upsilon\frac{d\Upsilon}{d\xi}=\frac{2\xi-\eta}{2\xi^{5/4}}\Upsilon+
\frac{(1-\eta\xi)(\eta-\nu^{-1}\xi^{-1})}{4\sqrt{\xi}}.
\label{jorma} \ee For $\chi\approx 0$, Eq.~(\ref{jorma}) is
approximated as \be
\Upsilon\frac{d\Upsilon}{d\zeta}+\frac{4}{\sqrt{3}}\Upsilon+\zeta\approx
0\,, \label{rien} \ee where \be
\zeta\!=\!-\frac{3\eta+2\xi}{2\sqrt{3}\,\xi^{\frac{1}{4}}}\!=\!
-\frac{3\chi+2}{2\sqrt{3}\,(\eta\chi)^{\frac{3}{4}}}\!=\!-\frac{1}{2\sqrt{3}}\frac{5+w_{\phi}}{1-w_{\phi}}
\Big|\frac{1-w_{\phi}}{1+w_{\phi}}\Big|^{\frac{3}{4}}.
\label{nada} \ee The solution of Eq.~(\ref{rien}) is \be
c_{1}|\Upsilon+\sqrt{3}\,\zeta|^3=c_{2}\Big|\Upsilon+\frac{1}{\sqrt{3}}\zeta\Big|\,,
\label{centro} \ee where $c_{1},c_{2}$ are integration constants
with $|c_{1}|+|c_{2}|>0$. In terms of the physical variables
$w_{\phi}, \Omega_{\phi}$, Eq.~(\ref{centro}) takes the form \bea
&&\!\!\!\!\!\!\!\!\frac{c_{1}}{|1+w_{\phi}|\sqrt{1-w_{\phi}^{2}}}\Bigg|\frac{1+w_{\phi}}{2}\pm\epsilon
\sqrt{2(\nu^{-1}+1)}\sqrt{\frac{1+w_{\phi}}{\Omega_{\phi}}}\Bigg|^3\nn\\
&&\!\!\!\!
=c_{2}\Bigg|\frac{7-w_{\phi}}{6}\mp\epsilon\sqrt{2(\nu^{-1}+1)}\sqrt{\frac{1+w_{\phi}}{\Omega_{\phi}}}\Bigg|,
 \label{river}
\eea which is the first integral between $w_{\phi},\Omega_{\phi}$
for $w_{\phi}\approx -1$. This equation represents the ``bottom''
parts of the lines $w_{\phi}(f)$ in Figure 2. Eq.~(\ref{river}) is
a cubic for $1/\sqrt{\Omega_{\phi}}$, and can, therefore, be
solved for $\Omega_{\phi}(w_{\phi})$. Then, from Eq.~(\ref{fla}),
(\ref{chiwphi}) one gets the expression $\psi(\chi)$, and plugging
into Eqs.~(\ref{find})-(\ref{ti}) all the relevant quantities can
be obtained, at least as quadratures.

\subsection{Special solutions}

For two particular values of $\nu$, we can immediately find from
Eq.~(\ref{dydx}) some special solutions of the system \bea &&
\nu=1/3\,\,\,\,:\,\,\,\,\psi=3/(1-3\chi) \label{special1}\\
&& \nu=9/7\,\,\,\,:\,\,\,\,\psi=3/(1-\chi)\,. \label{special2}
\eea The cases $\nu=1/3,\,9/7$ correspond to $\mu^{2}=9\beta/2,\,
81\beta/8$, for which values the general solution of the system we
discuss has been found in \cite{rubano} (for $V_{0}>0$), and
\cite{abel} respectively. For the solution (\ref{special1}) with
$V_{0}>0$ it arises that $\dot{\phi}>0$. Additionally, we can see
that the function $\chi(t)$ (and also $w_{\phi}(t)$) is
monotonically increasing with $0<\chi<1/3$, and that this solution
has an eternal acceleration for $\chi>(7-\sqrt{33})/24$.
Explicitly, for the solution (\ref{special1}) for both signs of
$V_{0}$ the various quantities are given in parametric form from
Eqs.~(\ref{find})-(\ref{q}) as follows \bea
\!\!\!\!\!\!\!&\phi&\!\!\!=\phi_{1}\!-\!\mu^{-1}\ln{(1-3\chi)}
=\phi_{1}\!-\!\mu^{-1}\ln\Big(2\frac{1+2w_{\phi}}{w_{\phi}-1}\Big) \label{phichi}\\
\!\!\!\!\!\!\!
&a^3&\!\!\!=\frac{9\rho_{0}e^{\mu\phi_{1}}}{V_{0}}\frac{\chi}{(1-3\chi)^2}=
\frac{9\rho_{0}e^{\mu\phi_{1}}}{4V_{0}}\frac{1-w_{\phi}^2}{(1+2w_{\phi})^2}\label{achi}\\
 \!\!\!\!\!\!\!&\Omega_{\phi}&\!\!\!=\frac{9\chi(1+\chi)}{(1+3\chi)^2}=
 \frac{9(1+w_{\phi})}{2(2+w_{\phi})^2}\\
\!\!\!\!\!\!\!&\rho_{\phi}&=-\frac{2V_{0}}{e^{\mu\phi_{1}}}\frac{1+2w_{\phi}}{(1-w_{\phi})^2}\label{rhophiwphi}\\
\!\!\!\!\!\!\!&p_{\phi}&
=\rho_{\phi}-\frac{2V_{0}}{e^{\mu\phi_{1}}}\Bigg(1\pm\sqrt{1-\frac{3e^{\mu\phi_{1}}}{2V_{0}}\rho_{\phi}}\Bigg)\label{foumara}\\
 \!\!\!\!\!\!\!&q&\!\!\!=\frac{1}{2}-\frac{27\chi(1-\chi)}{2(1+3\chi)^2}=
 \frac{1}{2}+\frac{27w_{\phi}(1+w_{\phi})}{4(2+w_{\phi})^2}\\
\!\!\!\!\!\!\!&t&\!\!\!\!=t_{1}\!\pm\!\frac{6e^{\frac{1}{2}\mu\phi_{1}}}{\sqrt{2}\mu
V_{0}}
\sqrt{\frac{V_{0}\chi}{1\!-\!3\chi}}=t_{1}\!\pm\!\frac{3e^{\frac{1}{2}\mu\phi_{1}}}{\mu
V_{0}}\sqrt{-\frac{V_{0}(1\!+\!w_{\phi})}{1\!+\!2w_{\phi}}}\nn\\
\!\!\!\!\!\!\!&\Leftrightarrow&\!w_{\phi}\!=\!-\frac{9e^{\mu\phi_{1}}+\mu^{2}V_{0}(t-t_{1})^{2}}
{9e^{\mu\phi_{1}}+2\mu^{2}V_{0}(t-t_{1})^{2}}\,,\label{tchi} \eea
where $\rho_{0}, \phi_{1}, t_{1}$ are integration constants. The
$\pm$ sign of Eq.~(\ref{foumara}) is independent of that appeared
throughout discerning the cases with $\dot{\phi}$ positive or
negative. For $V_{0}>0$, since $\rho_{\phi}>0$, from
Eq.~(\ref{rhophiwphi}) it is $w_{\phi}<-1/2$, and the relevant
sign in Eq.~(\ref{foumara}) is the negative one. In agreement with
the previous discussion, this solution starts decelerating with
$w_{\phi}\approx -1$, and as $w_{\phi}$ increases monotonically
with time, the universe accelerates resulting to $w_{\phi}=-1/2$.
Actually, since
$8q=31-3(8\Omega_{\phi}+3\sqrt{9-8\Omega_{\phi}})$, we have
acceleration for any $\Omega_{\phi}\gtrsim 0.37$. We can also see
that this solution possesses an event horizon \cite{suss}, since
$\int dt/a(t)\sim \chi^{1/6}\, _{2}F_{1}(1/6, 5/6, 7/6, 3 \chi)$,
and the limit for $\chi\!\rightarrow\! 1/3$ of this function is
finite.
\newline
The solution (\ref{special2}) can be easily seen that it does not
have accelerating intervals, so it is not of special interest
here.

\subsection{Late-times approximation}

Since equation (\ref{dydx}) contains phase-space variables, it is
suitable for approximating the general solution around the fixed
points of the system. We give for $V_{0}>0$ two more equivalent
forms of Eq.~(\ref{dydx}), which will be used to obtain the
general late-times evolution of the system. Defining \be
\tilde{\chi}=\frac{\varepsilon}{\chi-\nu}\,\,\,\,\,,\,\,\,\,\,Y=\chi^{-3/4}\,\psi^{-1}\,,
\label{ena} \ee Eq.~(\ref{dydx}) is written as \be Y\frac{dY}{d
\tilde{\chi}}=\frac{\varepsilon(2-\nu)\tilde{\chi}-1}{2\tilde{\chi}^{5/4}
(\nu\tilde{\chi}+\varepsilon)^{7/4}}Y+\frac{(1-\nu)\tilde{\chi}-\varepsilon}
{4\nu\tilde{\chi}^{3/2} (\nu\tilde{\chi}+\varepsilon)^{5/2}}\,,
\label{eqena} \ee where $\varepsilon=1$ (resp. $-1$) for
$\chi>\nu$ (resp. $\chi<\nu$).
 For
\be
\bar{\chi}=\frac{\varepsilon}{\chi-1}\,\,\,\,\,,\,\,\,\,\,Y=\chi^{-3/4}\,\psi^{-1}\,,
\label{dio} \ee Eq.~(\ref{dydx}) takes the form \be Y\frac{dY}{d
\bar{\chi}}=\frac{\varepsilon\bar{\chi}-1}{2\bar{\chi}^{5/4}
(\bar{\chi}+\varepsilon)^{7/4}}Y+\frac{(\nu-1)\bar{\chi}-\varepsilon}
{4\nu\bar{\chi}^{3/2} (\bar{\chi}+\varepsilon)^{5/2}}\,,
\label{eqdio} \ee where, as before, $\varepsilon=1$ (resp. $-1$)
for $\chi>1$ (resp. $\chi<1$).
\par
For $\mu^{2}<9\beta$, Eq.~(\ref{eqena}) is approximated close to
the fixed point (iii) as \be
Y\frac{dY}{du}+\frac{2-\nu}{\sqrt{1-\nu}}Y+ u\approx 0\,,
\label{psiapproxena} \ee where \bea \!\!\!\!u \!\!&=&\!\!
\frac{2\sqrt{1-\nu}}{3(2-\nu)}\,\left[\frac{3\nu+2}{\nu^{3/4}}-\frac{(3\nu+2)\tilde{\chi}+3\varepsilon}
{\tilde{\chi}^{1/4}(\nu\tilde{\chi}+\varepsilon)^{3/4}}\right]\\
\!\!&=&\!\!
\frac{2\sqrt{1-\nu}}{3(2-\nu)}\,\Big(\frac{3\nu+2}{\nu^{3/4}}-\frac{3\chi+2}{\chi^{3/4}}\Big)\label{year}\\
\!\!&=&\!\!
\frac{2\sqrt{1-\nu}}{3(2-\nu)}\,\Big(\frac{3\nu+2}{\nu^{3/4}}-\frac{3x^2+2y^2}{x^{3/2}\,\sqrt{y}}\Big)
 \label{u}\\
 \!\!&=&\!\!
 \frac{2\sqrt{1-\nu}}{3(2-\nu)}\,\Big(\frac{3\nu+2}{\nu^{3/4}}
-\frac{5+w_{\phi}}{(1+w_{\phi})^{3/4}(1-w_{\phi})^{1/4}}\Big),\label{uxux}
  \eea
while for $\mu^{2}>9\beta$, Eq.~(\ref{eqdio}) is approximated
close to the fixed point (iii) as \be
Y\frac{dY}{d\upsilon}+\sqrt{\frac{\nu}{\nu-1}}\,Y+\upsilon\approx
0\,, \label{psiapproxdio} \ee where \bea \!\!\!\!\upsilon
\!\!&=&\!\!
\frac{2}{3}\sqrt{1-\frac{1}{\nu}}\,\left[5-\frac{5\bar{\chi}+3\varepsilon}{\bar{\chi}^{1/4}(\bar{\chi}+
\varepsilon)^{3/4}}\right]\\
\!\!&=&\!\!
\frac{2}{3}\sqrt{1-\frac{1}{\nu}}\,\Big(5-\frac{3\chi+2}{\chi^{3/4}}\Big)\label{doc}\\
\!\!&=&\!\!
\frac{2}{3}\sqrt{1-\frac{1}{\nu}}\,\Big(5-\frac{3x^2+2y^2}{x^{3/2}\,\sqrt{y}}\Big)
 \label{upsilon}\\
 \!\!&=&\!\!
 \frac{2}{3}\sqrt{1-\frac{1}{\nu}}\,\Big(5-\frac{5+w_{\phi}}{(1+w_{\phi})^{3/4}(1-w_{\phi})^{1/4}}\Big)\,.
 \label{axax}
  \eea
\par
 For $\mu^{2}<9\beta$, the solution of Eq.~(\ref{psiapproxena}) is
 \be
c_{1}\left|Y+\sqrt{1-\nu}\,
u\right|^{1-\nu}=c_{2}\Big|Y+\frac{1}{\sqrt{1-\nu}}u\Big|\,,
 \label{solution1}
 \ee
 where $c_{1}, c_{2}$ are integration constants with
 $|c_{1}|+|c_{2}|>0$. Concerning Eq.~(\ref{psiapproxdio}), its solution for $9\beta <\mu^{2}< 72\beta /
 7$ is
 \be
c_{3}\,|Y+(\sigma-\tau)\upsilon|^{\sigma-\tau}=c_{4}\,|Y+(\sigma+\tau)
\upsilon|^{\sigma+\tau}\,,
 \label{solution2}
 \ee
while, for $\mu^{2}>72\beta/7$ the solution is \be
\ln\left|Y^{2}+2\sigma\upsilon
Y+\upsilon^{2}\right|-\frac{2\sigma}{\sqrt{1-\sigma^{2}}}\arctan
\frac{Y+\sigma\upsilon}{\sqrt{1-\sigma^{2}}\,\upsilon}=c\,.
\label{solution3} \ee In the above two equations, $c_{3}, c_{4},
c$ are integration constants with
 $|c_{3}|+|c_{4}|>0$, and $\sigma=1/2\sqrt{1-\nu^{-1}}$, $\tau=\sqrt{(4-3\nu)/4(\nu-1)}$.
\newline
We can write $Y$ from Eq.~(\ref{pop}) in terms of $x,y$ as \be
Y=\Big(\frac{y}{x}\Big)^{3/2}\,\Big[1+\Big(x-\sqrt{\nu^{-1}+1}\Big)\frac{x}{y^{2}}\Big]\,,
\label{the} \ee and thus, replacing $Y, u, \upsilon$ in terms of
$x,y$ from
 Eq.~(\ref{the}), (\ref{u}), (\ref{upsilon}) respectively,
 {\it{Eqs.~(\ref{solution1}), (\ref{solution2}), (\ref{solution3})
become late-times first integrals of the system in the plane
$(\dot{\phi}/H, \sqrt{V}/H)$, approaching respectively, for
$\mu^{2}<9\beta$ a scalar field dominated attractor (node), for
$9\beta<\mu^{2}<72\beta/7$ a node ``scaling'' attractor, and for
$\mu^{2}>72\beta/7$ a spiral ``scaling'' attractor}}.
\newline
Furthermore, we can write $Y$ from Eq.~(\ref{fla}), (\ref{q}) in
terms of $w_{\phi},\Omega_{\phi}$ or $w_{\phi},q$ as \bea
\!\!\!\!\!\!Y\!\!&=&\!\!\frac{\sqrt{2}}{(1-w_{\phi}^{2})^{1/4}}\Bigg(\sqrt{\frac{2}{1+w_{\phi}}}-
\sqrt{\frac{\nu^{-1}+1}{\Omega_{\phi}}}\Bigg)\label{oxox}
\\
\!\!&=&\!\!\frac{\sqrt{2}}{(1-w_{\phi}^{2})^{1/4}}\Bigg(\sqrt{\frac{2}{1+w_{\phi}}}-
\sqrt{\frac{3(\nu^{-1}+1)w_{\phi}}{2q-1}}\Bigg)\,.\label{lou}\eea
Thus, replacing $Y,u,\upsilon$ in terms of
$w_{\phi},\Omega_{\phi}$ from Eqs.~(\ref{oxox}), (\ref{uxux}),
(\ref{axax}) respectively, {\it{ Eqs.~(\ref{solution1}),
(\ref{solution2}), (\ref{solution3}) become late-times first
integrals of the system between the variables $w_{\phi},
\Omega_{\phi}$}}.
\newline
Finally, $\Omega_{\phi}$ can be written from its definition as
$\Omega_{\phi}=1-\beta\rho_{0} (1+z)^{3}/H^{2}$, and thus, $Y$ is
written as \be
Y\!=\!\frac{\sqrt{2}}{(1-w_{\phi}^{2})^{1/4}}\Bigg(\sqrt{\frac{2}{1\!+\!w_{\phi}}}-
\sqrt{\frac{\nu^{-1}\!+\!1}{1\!-\!\beta\rho_{0}(1+z)^{3}H^{-2}}}\Bigg).\label{lobi}
 \ee Replacing $Y$ from Eq.~(\ref{lobi}), {\it{Eqs.~(\ref{solution1}), (\ref{solution2}), (\ref{solution3})
become late-times first integrals of the system between the
variables $w_{\phi},(1+z)^{3}/H^{2}$}}.

\section{Conclusions}

In conclusion, we have studied how long a scalar field of
exponential potential can play the role of quintessence in a
realistic universe filled with pressureless baryonic matter. By
deriving and studying directly the first order differential
equation of the deceleration parameter $q$ with respect to the
state parameter of the scalar field $w_{\phi}$, we found that
there exist whole classes of solutions which enter into
accelerating eras not at late times, independently of the
parameter $\mu$ determining the steepness of the potential.
Moreover, the condition $w_{\phi}<-1/3$ does not necessarily imply
acceleration. Observational constraints of $\Omega_{m0}, w_{\phi
0}$ provide immediate bounds on $\mu$ for the viable solutions,
irrespectively of initial conditions or other parameters, in
accordance -- as it has been claimed \cite{kolda} -- with all the
current observations. Using the same first order equation, we have
obtained the solution $\Omega_{\phi}(w_{\phi})$ for the almost
cosmological constant case $w_{\phi}\approx -1$, and the generic
late-times solution of the system in the plane of the variables
$(\dot{\phi}/H,\sqrt{V}/H)$, $(w_{\phi},\Omega_{\phi})$,
$(w_{\phi},q)$, and $(w_{\phi},(1+z)^{3}/H^{2})$.

\[ \]
{\bf Acknowlegements} We wish to thank J. Garriga, E. Kiritsis,
and R. Maartens for useful discussions. This work is partially
supported by the European RTN networks HPRN-CT-2000-00122 and
HPRN-CT-2000-00131.  G.K. acknowledges partial support from
FONDECYT grant 3020031 and from Empresas CMPC. Centro de Estudios
Cient\'{\i}ficos is a Millennium Science Institute and is funded
in part by grants from Fundacion Andes and the Tinker Foundation.

\end{document}